\documentclass{aa}
\usepackage{hyperref}           
\hypersetup{colorlinks,linkcolor={blue},citecolor={Green},urlcolor={blue}}
\usepackage[dvipsnames]{xcolor}
\usepackage{txfonts}
\usepackage{times}
\usepackage{lscape}

\begin{document}


    \title{Relationship between TIGRE solar S-index and USET Ca {\sc ii} K full disk images}
    \author{G. Vanden Broeck\inst{1,2} \and S. Bechet\inst{1} \and F. Clette\inst{1} \and G. Rauw\inst{2} \and K.-P. Schröder\inst{3} \and M. Mittag\inst{4}}

    \institute{Department of Solar Physics and Space Weather, Royal Observatory of Belgium (ROB), Av. Circulaire 3, 1180 Uccle, Belgium \\ email: \href{mailto:gregory.vandenbroeck@observatory.be}{gregory.vandenbroeck@observatory.be}
    \and University of Liège, Allée du 6 août, 19c - Bât. B5c, B-4000 Liège (Sart-Tilman), Belgium
    \and Departamento de Astronomía, Universidad de Guanajuato, Apartado Postal 144, 36000 Guanajuato, Mexico
    \and Hamburger Sternwarte, Universität Hamburg, Gojenbergsweg 112, 21029 Hamburg, Germany}


    \abstract
    {Full disk observations of the solar chromosphere in the Ca {\sc ii} K line represent a valuable dataset for studies of solar magnetic activity. The well known S-index is widely used to investigate the magnetic activity of stars, however,  its connection to the coverage of stellar magnetic structure is still poorly understood.}
    {We use the archives of full disk Ca {\sc ii} K images taken by the Royal Observatory of Belgium with the Uccle Solar Equatorial Table (USET) to derive the area fraction of the brightest chromospheric structures over the last decade. These data have allowed us to study the end of the solar cycle 24 and the beginning of the solar cycle 25. Our aim is to compare this dataset with the solar S-index from the Telescopio Internacional de Guanajuato Robotico Espectroscopico (TIGRE) lunar spectroscopy to analyze the relationship between a disk coverage index and an integrated spectrum. We also searched for periodic modulations in our two datasets to detect the solar rotation period.}
    {We used more than 2700 days of observations since the beginning of the Ca {\sc ii} K observations with USET  in July 2012. We performed a calibration of the images (re-centering and center-to-limb variation correction).  The brightest regions of the solar surface (plages and enhanced network) were then segmented using an algorithm based on an intensity threshold. We computed the area fraction over the solar disk and compared it with the S-index from TIGRE. For the detection of periodic modulations, we applied a discrete Fourier power spectrum method to both datasets.}
    {A tight linear relationship was found between the USET area fraction and the TIGRE S-index, with an improved correlation obtained in the low-activity regime by considering the enhanced network. In both time series, we detected the modulation caused by the rotation of bright structures on the solar disk. However, this detection is constrained in the case of TIGRE due to its observation strategy.}
    {We studied the correlation between the disk coverage with chromospheric structures and the variability of the S-index on an overlapping period of ten years. We concluded that the disk coverage index is a good proxy for the S-index and will be useful in future studies of the magnetic activity of solar-type stars. 
    The USET area fraction dataset is most appropriate for evaluating the solar rotation period and will be used in future works to analyze the impact of the inclination of the stellar rotation axis on the detectability of such periodic modulations in solar-type stars.}

    \keywords{Sun: activity -- Sun: chromosphere -- Sun: faculae, plages -- Star: activity -- Star: solar-type}
    \maketitle
%

\section{Introduction}\label{sec:Introduction}

    Since 1960, the chromospheric activity of a large number of stars has been monitored in the Ca {\sc ii} K and H lines. This program was initiated in the context of the HK  project at Mount Wilson \citep{Wilson-1978} and eventually led to the creation of the Mount Wilson S-index, further denoted as $S_{\textrm{MWO}}$. Over more than three decades, hundreds of stars (including the Sun) were observed to search for stellar activity \citep{Baliunas-1998}. The HK project was followed by the Lowell's monitoring, a complementary synoptic survey at the Lowell observatory with the Solar-Stellar Spectrograph (SSS) \citep{Hall-2007} and the TIGRE project \citep{Schmitt-2014}. Additional stellar activity surveys come from radial velocity exoplanet searches where S-indices can be derived as a by-product, such as HARPS \citep{Lovis-2011} and California Planet Search \citep{Isaacson-2010}. Recently the arrival of large spectroscopic surveys \citep{Zhang-2022} has allowed for observations of hundreds of thousands of stars to be carried out.

    The Ca {\sc ii} K line is also used to monitor the Sun, the only star for which we can resolve the surface in detail. The observations of the Sun's full disk in this line started in 1893 in Meudon and are still running in various sites around the world, supplying a long-term dataset \citep{2022-Chatzistergos}. The images show the lower chromosphere and provide information on plages and network regions that are the manifestations of surface magnetic fields in this part of the solar atmosphere.  Thus, they can be used to study the evolution of magnetic activity.

     While the S-index is widely used to study the magnetic activity of late-type stars, its connection to the coverage of stellar magnetic structure is poorly understood \citep{Shapiro-2014, 2018-Meunier, Dineva-2022, 2023-Singh}; in particular, the effect of the inclination of the stellar rotation axis on the S-index variability \citep{Shapiro-2014, Sowmya-2021}. Our goal in this article is to study the relationship between the disk coverage with chromospheric structures and the variability of the solar S-index to provide a useful proxy for subsequent studies on solar-type stars. 
    
    To assist in approaching an answer to these questions, we have taken advantage of two datasets taken simultaneously over the past decade: the solar S-index from TIGRE and the Ca {\sc ii} K full disk images from USET. In Section \ref{sec:Dataset}, we present the technical specifications of those datasets. In Section \ref{sec:indices}, we describe the image processing method to construct a disk-resolved index from full disk images. In Section \ref{sec:Fourier_analysis} we study the modulation produced by the bright structures on the time series. Finally, in Section \ref{sec:USET_vs_TIGRE}, we study the correlation between the two indices and derive a proxy for the S-index based on Ca {\sc ii} K images. This proxy will be used in a subsequent paper to study the magnetic activity of solar-type stars and the effect of the inclination of the rotation axis.

\section{Datasets}\label{sec:Dataset}

    \subsection{USET}\label{subsec:USET_data}

\begin{figure}[t]
            \centering
            \includegraphics[width=\linewidth]{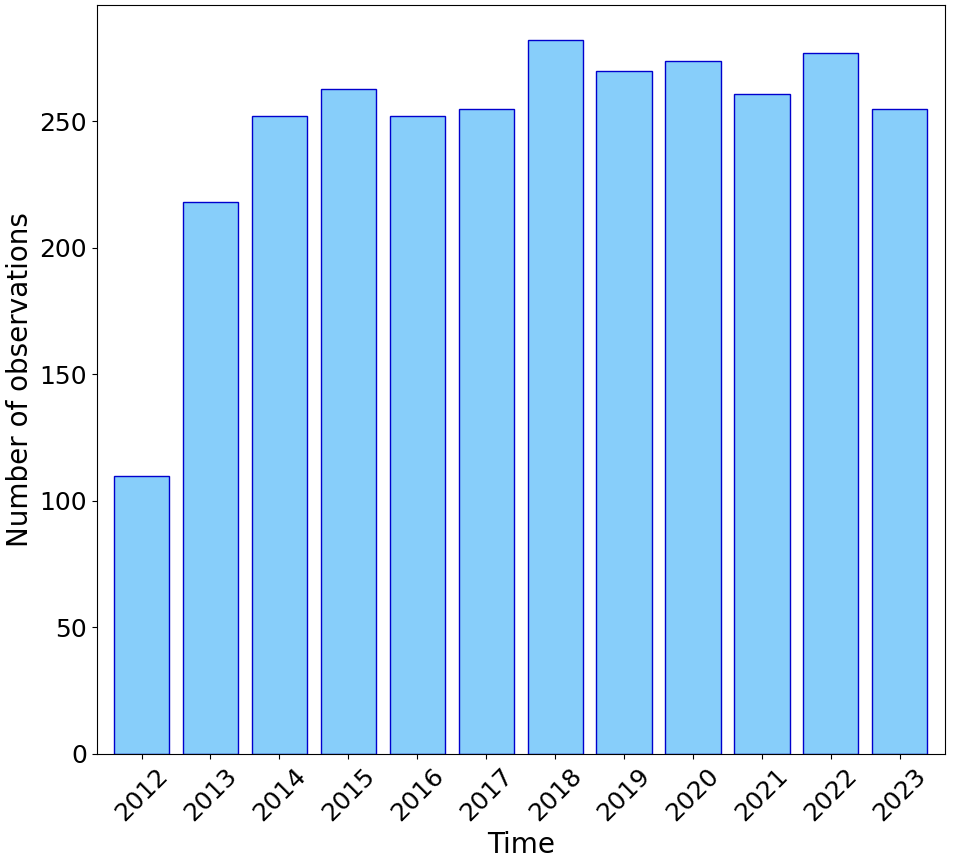}
            \caption{Number of Ca {\sc ii} K images per year from the USET station.}
            \label{fig:USET_Ca_observations}  
        \end{figure}

        The Uccle Solar Equatorial Table (USET) station at the Royal Observatory of Belgium (ROB), located in Uccle, south of Brussels, has been acquiring full disk solar images in the Ca {\sc ii} K line since July 2012 \citep{2023-USET}. In addition, the station carries three other solar telescopes (white-light, 656.3 H-alpha, and sunspot drawings) to  simultaneously monitor the photosphere and the chromosphere. Figure \ref{fig:USET_Ca_observations} shows the total number of days of observations per year, with an average of 260 per year. The gaps are essentially due to bad weather.
        
        The optical set-up consists in a refractor of 925 mm focal length and 132 mm aperture. The filter is temperature-controlled and its central wavelength is  $\lambda$ = 3933.67$\AA$, with a bandwidth of 2.7$\AA$. The images are acquired with a 2048x2048 CCD with a dynamic range of 12 bits. The instrumental set-up is the same since 2012, except for the introduction of an additional neutral filter on July 10, 2013. The acquisition cadence can go up to 4 frames per second in case of transient events to record, and the daily synoptic cadence is 15 minutes.
        
    \subsection{TIGRE }\label{subsec:TIGRE_data}
    
        The Telescopio Internacional de Guanajuato Robotico Espectroscopico (TIGRE) is a 1.2 m robotic telescope located in Guanajuato, central Mexico \citep{Schmitt-2014}. With its sole instrument HEROS, a spectrograph with a spectral resolution R $\sim$ 20.000, TIGRE has collected more than 48.000 spectra of 1.151 different sources \citep{Gonzalez-Perez-2022}. To obtain an integrated solar S-index, TIGRE observes the light reflected by the Moon. Therefore, it depends on the lunar phases and observations are interrupted for a few days around the New Moon. TIGRE has observed the Sun-as-a-star continuously since August 2013, except for a few gaps due to instrumental problems, with a total of approximately 1200 days.

\section{Assessment of indices}\label{sec:indices}
    
    \subsection{USET disk-resolved index}\label{subsec:USET_index}
        The first step of the image processing consists of the automatic limb fitting, based on a combination of Canny edge detection and morphological operations \citep{2009-Gonzales}. Using the limb estimation, we detected the disk center and calculate the radius, assuming that the shape of the solar image is a circle. Finally, the disk was recentered in the middle of the image and the meta-data were filled following the SOLARNET standard \citep{solarnet-2020}, making level 1 of the image series. For this analysis, an automatic quality selection was performed to keep the best image recorded per day. Images with opaque clouds hiding some parts of the solar disk or strong atmospheric turbulence have been discarded. The non-radial inhomogeneity due to the presence of high-altitude veils or transparent clouds has not been corrected. This reduces the final sample to 2725 images.
    
        The solar images showing the chromosphere present an intensity variation from the center to the limb, the so-called center-to-limb variation (CLV). This effect must be corrected to have the same level of background intensity relative to the chromospheric emissions that have been identified; otherwise, a different threshold should be applied depending on the considered region. The steps of the correction algorithm are the following.
        
        \begin{enumerate}
            \item We performed a polynomial fit on the order of 5 for a set of angles between 0$^{\circ}$ and 360$^{\circ}$, with a step of 1$^{\circ}$. For every angle, we calculate the residuals between the intensity profile along the radius and the fit and we kept the fit for the angle where the residuals are minimum. This gives a first guess of the intensity as a function of the radius without the presence of sunspots or plages.
            \item We created a mask of the quiet solar disk $I_{QS}$ based on that fit.
            \item A first-guess corrected image was then obtained by $I_C = I_i/I_{QS}$ where $I_i$ is the intensity of the initial image and $I_C$ is the intensity of the corrected image.
            \item Based on this first guess, we removed the bright plages following the segmentation method described in \cite{Chatzistergos-2019} so that the next step does not use those bright structures in the computation.
            \item Finally, we repeated  steps 1, 2, and 3 but with step 1 (fit of the intensity profile) performed by computing the mean intensity of the whole image without considering the bright plages.
        \end{enumerate}
    
        \begin{figure}[t]
            \centering
            \includegraphics[width=\linewidth]{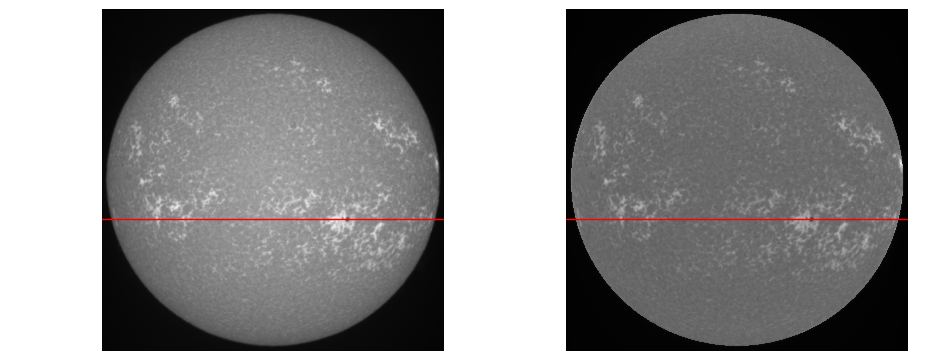}
            \includegraphics[width=\linewidth]{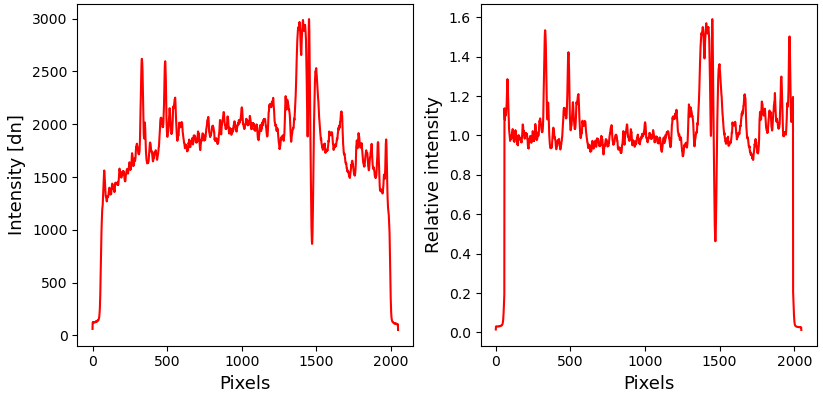}
            \caption[CLV]{Example of the results after the CLV correction. Top panels display the Ca {\sc ii} K images from the 16th of January 2015 before the correction (left) and after the correction (right) while bottom panels display the intensity variation on the horizontal red line, from the left to the right of each image. The intensities are expressed in term of digital number (dn) for the raw image and intensity relative to the quiet Sun regions for the corrected image.}
            \label{fig:CLV_correction}  
        \end{figure}
       
        Figure \ref{fig:CLV_correction} shows an example of the intensity variations across the solar disk before and after applying the CLV correction. On the corrected image (right image), the pixel intensity is normalized to 1 (quiet Sun regions) with minimal values at 0 (out of the solar disk). The variations show peaks upwards and downwards, which are due to the presence of plages and sunspots respectively. The small intensity variations are representative of the chromospheric network intensity.
                       
\begin{figure}[t]
            \centering
            \includegraphics[width=0.4\hsize]{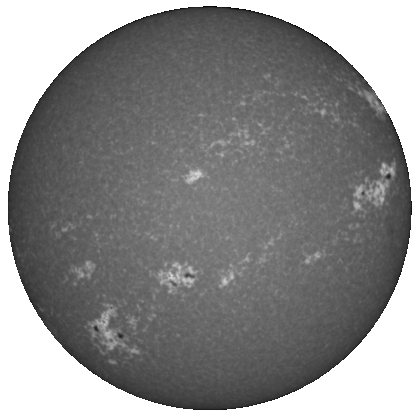} \\
            \includegraphics[width=0.4\hsize]{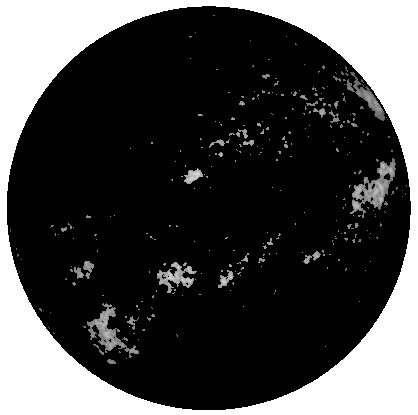}
            \includegraphics[width=0.4\hsize]{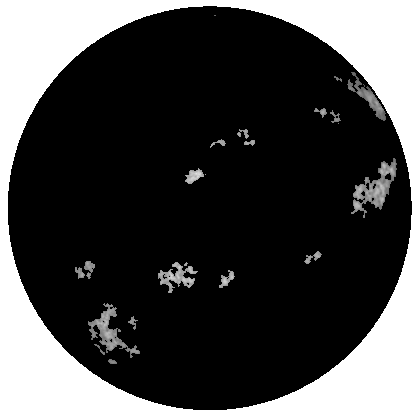}
            \caption{Example of the results from the segmentation process. Top image represents the recentered raw image from the 29th of October 2013. Bottom images are the results of the segmentation with and without the enhanced network. Left image: plages and enhanced network; Right image shows the plages only.}
            \label{fig:Plages_segmentation}
        \end{figure}

        The segmentation algorithm follows the method described in \cite{Chatzistergos-2019}. This method assumes a Gaussian brightness distribution of the pixels in the image, dominated by inactive areas known as the so-called quiet Sun. This distribution is enhanced on either side due to spots and bright structures. Due to their small spatial coverage, sunspots have a small contribution on the low side of the distribution. On the other hand, bright and extended structures have a more significant contribution and enhance the distribution for high intensity values. In this study, we segmented the plages (P), which are the chromospheric counterparts of the faculae. These structures have been extensively used to study the chromospheric activity. Here, in addition, we considered the enhanced chromospheric network (EN). This has been defined in \cite{Worden-1998} and \cite{2023-Singh} as small regions of decaying plages, dispersing into bright patches. The enhanced network is much smaller than the plages but just as bright. The segmentation of both structures is based on the same intensity threshold estimated iteratively from the intensity distribution and we distinguish the plages and the enhanced network based on an area threshold. At the center of the disk, the value of the area fraction threshold is 0.0004 and it decreases with distance from the center (taking the deprojection of the area into account). To prevent artifacts resulting from processing the last few pixels near the limb, our segmentation algorithm exclusively takes pixels within 99\% of the solar disk radius into account. Figure \ref{fig:Plages_segmentation} shows an example of the results of the segmentation process for the plages with the enhanced network and for the plages only. As expected, the enhanced network can be observed as bright patches that are close to the extended bright structures, but not covering the entire solar surface. The impact of the enhanced network will be discussed further in Section \ref{subsec:correlation_tigre_uset}.
    
\begin{figure}[t]
            \centering
            \includegraphics[width=\hsize]{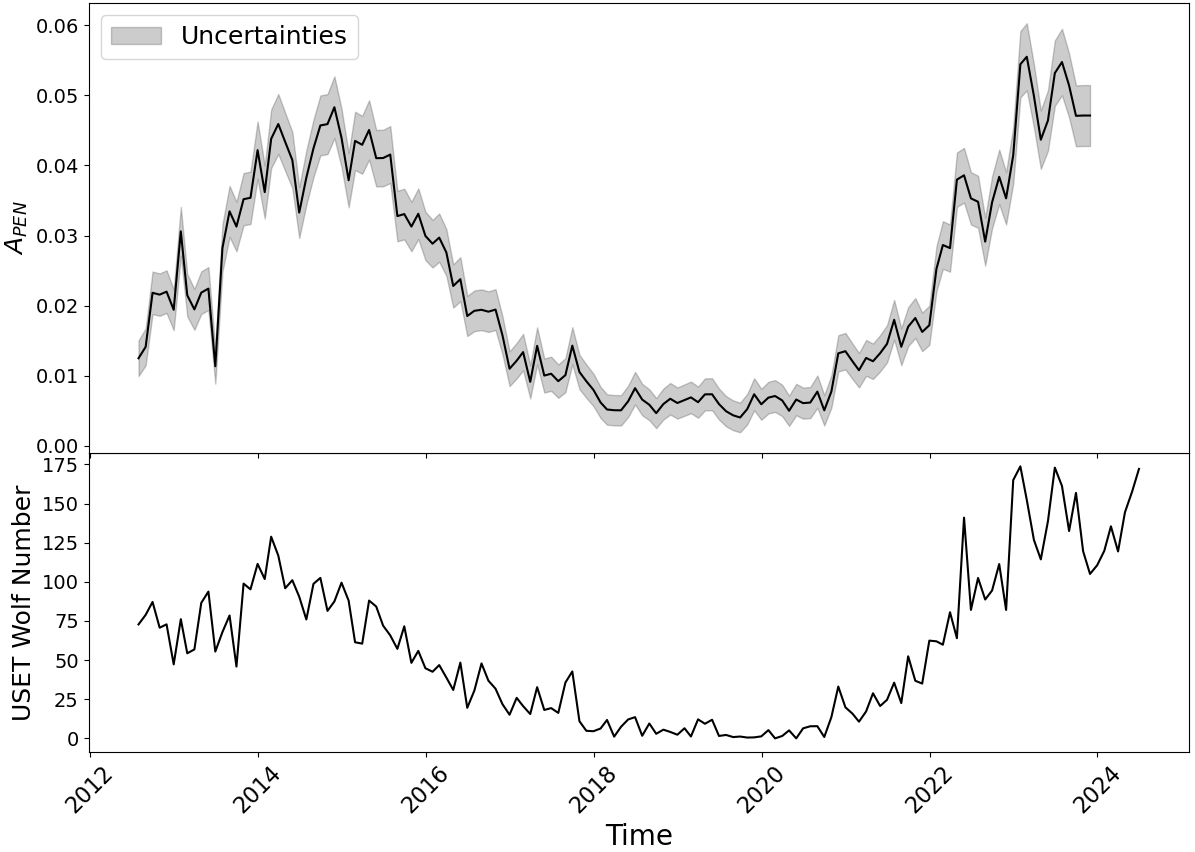}
            \caption{Temporal evolution of monthly averaged $A_{PEN}$, the area of plages, and the enhanced network, expressed as  fractions of the solar disk (top). Temporal evolution of monthly averaged USET Wolf number (bottom). Uncertainties on the $A_{PEN}$ depend linearly on the area.}
            \label{fig:Evolution_Areafrac_SN}
        \end{figure}

        The area of plages and enhanced network, $A_{PEN}$, expressed in fraction of solar disk, is the number of pixels of plages and enhanced network over the total number of solar disk pixels. Figure \ref{fig:Evolution_Areafrac_SN} illustrates the evolution of the monthly averaged $A_{PEN}$ as a function of time, with the uncertainty increasing with the area. The error bars on $A_{PEN}$ were estimated from dates where more than four images not affected by clouds were available. For such days, we computed the mean and the standard deviation of $A_{PEN}$. After removing some outliers, deviating by more than 3$\sigma$, we found that the error follows a roughly linear trend with $A_{PEN}$ and, thus, we adjusted a linear relation to compute the uncertainty for any value of $A_{PEN}$. The plot shows a long-term modulation following the solar cycle, as illustrated in the bottom panel with the Wolf number evolution from the USET sunspot drawings. The Wolf number is based on straight counts separating groups and individual spots. It reflects the emergence of magnetic flux \citep{Clette-2016}. Shorter modulations are also present and are due to the solar rotation. This is studied further in Section \ref{sec:Fourier_analysis}.
                    
    \subsection{TIGRE S-index}\label{subsec:TIGRE_index}

        The Mount Wilson S-index quantifies the chromospheric emission in the core of the Ca {\sc ii} H \& K lines, measuring the flux in a triangular shape bandpass with a full width at half maximum (FWHM) of 1.09 $\AA$, denoted by $N_H$ and $N_K$ in Eq. (\ref{eq:S_MWO_formula}). To minimize the impact of atmospheric turbulence and varying atmospheric emissions, the measured line core fluxes are normalized relative to the flux of two bandpasses of 20$\AA$ width in nearby continua redwards and bluewards of the H \& K lines, denoted by $N_R$ and $N_V$, respectively. A multiplicative factor $\alpha$ is introduced to standardize different instruments and calibrate them on the same S-index scale \citep{Vaughan-1978, Duncan-1991}, so that the $S_{\textrm{MWO}}$ is defined as:
    
        \begin{equation}
            S_{\textrm{MWO}} = \alpha \left(\frac{N_H + N_K}{N_R + N_V} \right)
            \label{eq:S_MWO_formula}
        .\end{equation}
        
        \begin{figure}[t]
            \centering
            \includegraphics[width=\linewidth]{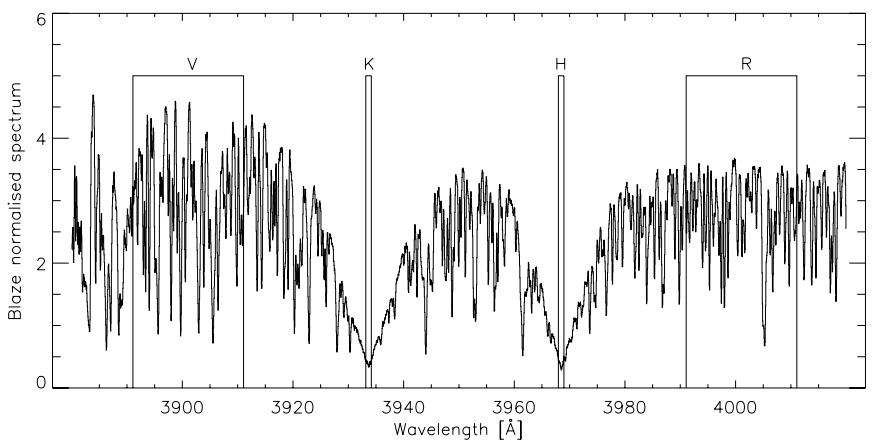}
            \caption{Solar Ca {\sc ii} H \& K spectrum as inferred from lunar spectra along with the TIGRE bandpasses used for the S-index calculation \citep{Mittag-2016}. The bandwidth is 1$\AA$ in the core of the lines and about 20$\AA$ in the two nearby continua. Bandpasses in the core of the lines are rectangular.}
            \label{fig:S-index-spectrum}  
        \end{figure}

        The S-index measured by TIGRE, denoted as $S_{\textrm{TIGRE}}$, represents the emission in the core of the Ca {\sc ii} H \& K lines with a rectangular bandpass of 1$\AA$ (see Figure \ref{fig:S-index-spectrum}), instead of a triangular bandpass as with Mt. Wilson. In order to compare the TIGRE results with the original Mt. Wilson measurements, the $S_{\textrm{TIGRE}}$ values are converted to the Mt. Wilson system by means of regular observations of a sample of 50 stars with very well known $S_{\textrm{MWO}}$ values (see Figure 2 of \cite{Mittag-2016} for the results of this comparison). A well-defined linear transformation was found:
        
        \begin{equation} 
            \label{eq:S_tigre_equation}
            S_{\textrm{MWO}} = (0.0360 \pm 0.0029) + (20.02 \pm 0.42) \ S_{\textrm{TIGRE}}
        .\end{equation}
        
\begin{figure}[t]
            \centering
            \includegraphics[width=\hsize]{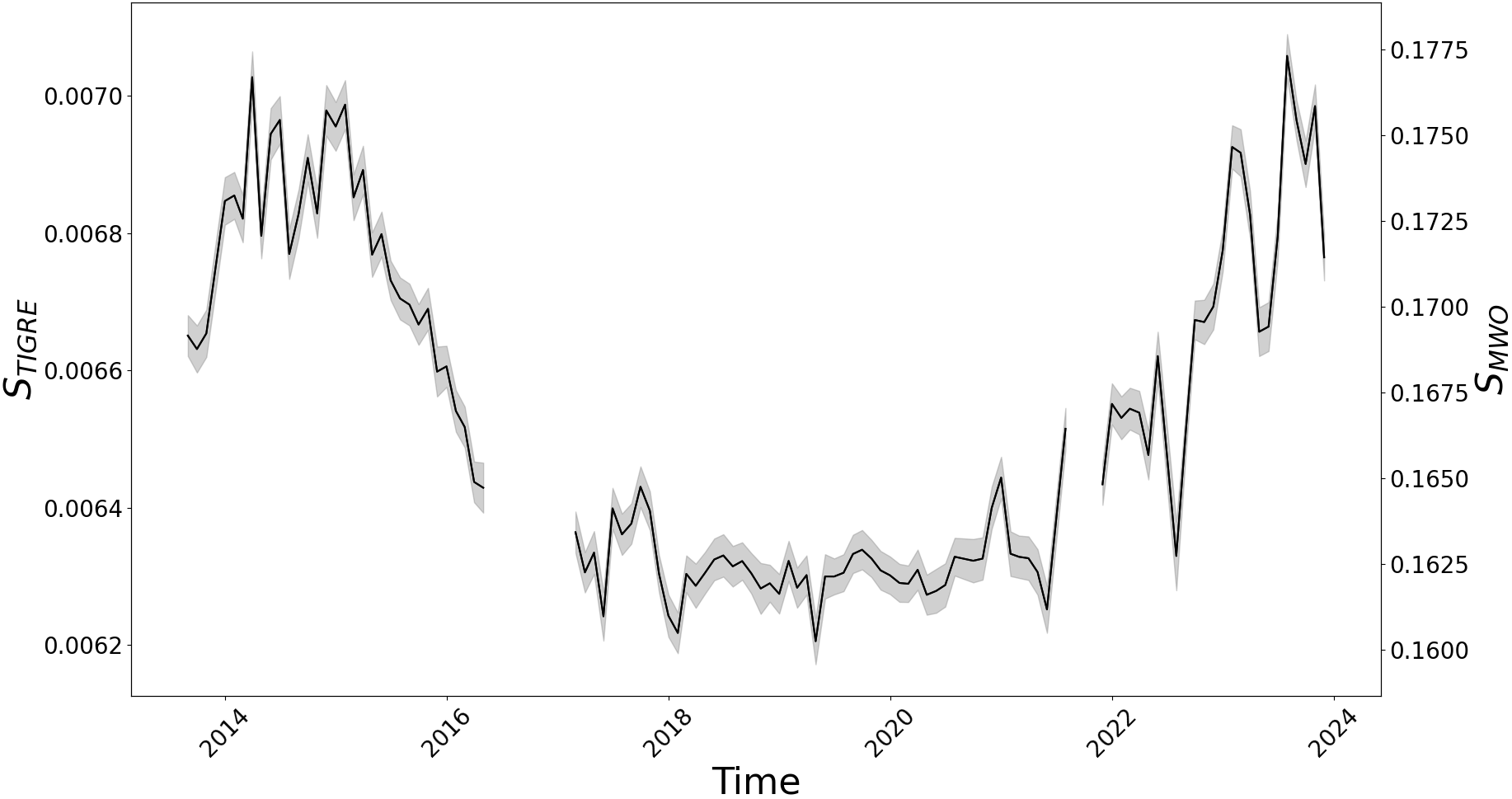}
            \caption{Temporal evolution of the monthly averaged solar S-index from TIGRE covering almost a complete solar cycle (from August 2013 to December 2023). Left axis displays the $S_{\textrm{TIGRE}}$ values while the right axis shows them in the Mt. Wilson scale using Eq. (\ref{eq:S_tigre_equation}).}
            \label{fig:S_index_evolution}
        \end{figure}

        Hereafter, we use Eq. (\ref{eq:S_tigre_equation})  to convert TIGRE measurements into the $S_{\textrm{MWO}}$ index throughout the rest of the paper. Figure \ref{fig:S_index_evolution} shows the temporal evolution of the monthly averaged solar S-index from TIGRE for both scales: $S_{\textrm{TIGRE}}$ (left) and $S_{\textrm{MWO}}$ (right). The exposure time of the spectra were carefully chosen to reach a typical uncertainty of 1\% in $S_{\textrm{TIGRE}}$, including potential adjustments to compensate for bad seeing conditions or elevated extinction \citep{Hempelmann-2016}. The TIGRE to Mt. Wilson S-index calibration uses the same 40 calibration stars as the Mount Wilson team \citep{Baliunas-1995}; hence, this guarantees a good long-term stability. Any drifts would require a larger fraction of those calibration stars to change their activity in the same sense. However, these were selected by Olin Wilson over a long period of time as relatively invariable and the size of the calibration star sample should be sufficient to average out any individual trends. The gaps in Figure \ref{fig:S_index_evolution} refer to the two main instrumental issues that TIGRE had faced: in 2016, the observations were stopped for $\sim$ 9 months due to mirror cell refurbishment and mirror aluminization; in 2021, the uninterruptible power supply batteries were in poor conditions and prevented the observations for $\sim$ 3 months \citep{Gonzalez-Perez-2022}. Nevertheless, the long-term variation due to the solar activity cycle is evident and we also observe some variations on the short-term scale. This is discussed further in the next section.
        
\section{Solar cycle and solar rotation modulations from USET and TIGRE}\label{sec:Fourier_analysis}

    \subsection{USET disk-resolved index}\label{subsec:USET_index_modulation}
    
        To search for the presence of a rotational modulation in the time series of $A_{PEN}$, we used the discrete Fourier power spectrum method developed by \cite{Heck-1985} amended by \cite{Gosset-2001}. This method explicitly accounts for the irregularities that typically affect astronomical time series creating gaps due to poor weather conditions or instrumental problems. The Fourier method offers a sensitive tool to search for the presence of periodic modulations: the existence of such a periodic signal leads to a peak in the power spectrum at the corresponding frequency.
        
        \begin{figure}[t]
            \centering
            \includegraphics[width=0.49\hsize]{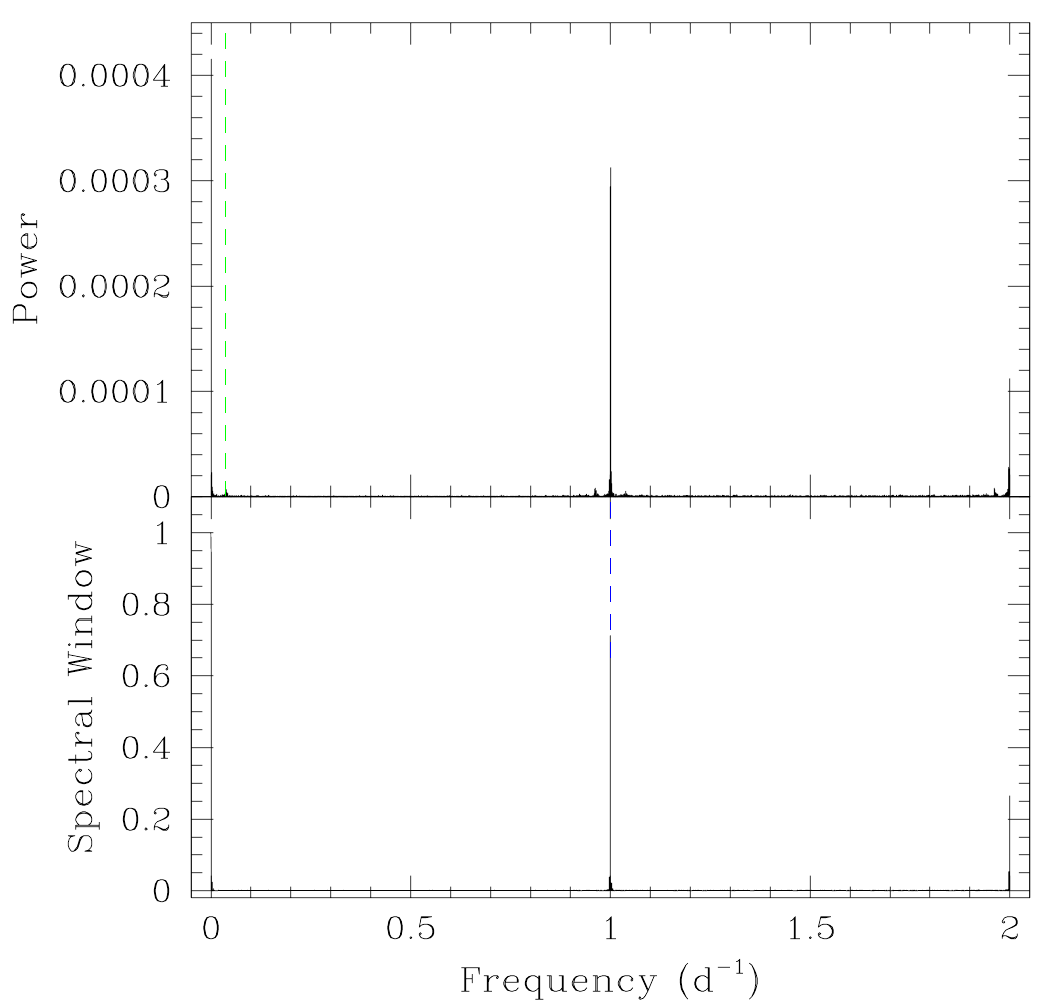}
            \includegraphics[width=0.49\hsize]{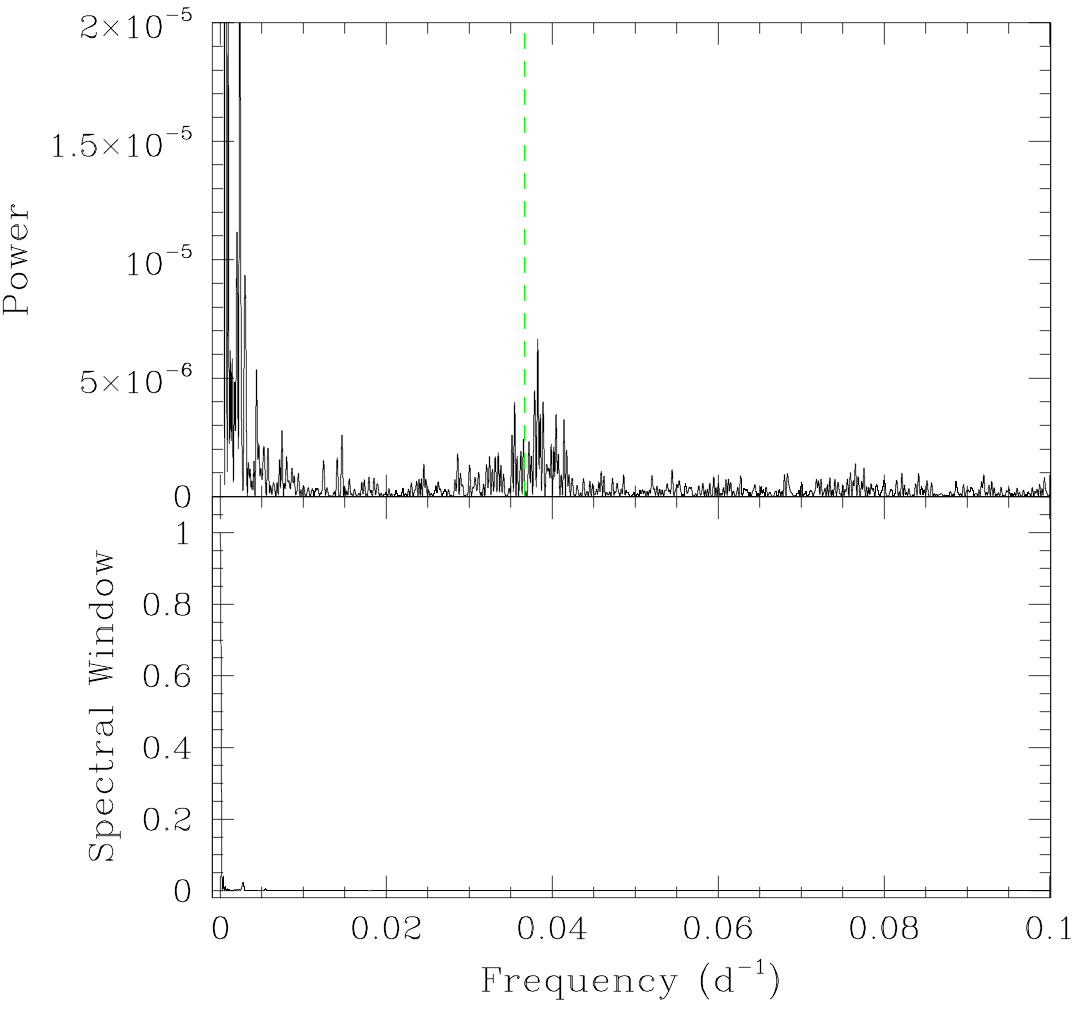}
            \caption{ Fourier power spectrum (top) and spectral window (bottom) for the $A_{PEN}$ time series for frequencies between 0 and 2 d$^{-1}$ (left). The green dashed line in the power spectrum panel yields $\nu_{\rm Car}$, while the blue dashed line in the spectral window identifies the main aliasing frequency at 1 d$^{-1}$. Right: Same, but zooming on the frequency range between 0 and 0.1 d$^{-1}$. \label{fig:USET_power_spectrum}}
        \end{figure}

        Figure \ref{fig:USET_power_spectrum} illustrates the power spectrum of the full time series of $A_{PEN}$ for frequencies between 0 and 2 d$^{-1}$, as well as a zoom on the region between 0 and 0.1 d$^{-1}$. The lower panel displays the spectral window which illustrates the aliasing phenomenon that results from the sampling of the time series. Because the USET data are taken with a nearly regular sampling at the pace of one observation per mean solar day, except for the gaps due to bad weather, the spectral window contains only peaks at integer multiples of 1 d$^{-1}$. For a time series containing a signal of true frequency, $\nu,$ sampled with a regular time step of 1 day, the power spectrum will not only host a peak at $\nu$, but also at the aliasing frequencies 1 + $\nu$, 1 - $\nu$, and so on.

        The power spectrum of the $A_{PEN}$ full time series is dominated by a peak at low frequency associated with the long-term variability arising from the solar activity cycle. The highest peak is however not located exactly at the frequency of the 11-year cycle (around 0.00025 d$^{-1}$). This is no surprise since the USET data cover the period from July 2012 until November 2023, which is only about a single solar activity cycle (i.e., large parts of cycle 24 and the rising part of cycle 25). The second important feature in the power spectrum is a group of peak which are close, although not strictly identical, to the frequency ($\nu_{\rm Car} = 0.0367$ d$^{-1}$) associated with the Carrington synodic rotation period (27.2753\,d). The highest power is recorded at a frequency of 0.0383 d$^{-1}$ (period of 26.1342 d, close to the synodic rotation period of the Sun at the equator), although as we will see below, the exact location of the highest peak changes with epoch and the signal is actually the manifestation of a quasi-periodic rather than a genuine periodic phenomenon. These peaks reflect rotational modulation resulting from an asymmetry in the longitudinal distribution of active regions. An important point to note is that the group of peaks in the power spectrum extends over a range in frequencies that is significantly (a factor of 10) broader than the natural width expected from the total duration of the time series. This reflects the fact that the plages are spread over a range of latitudes hence are modulated by a range of rotational periods. Therefore, the detected group of peaks is a blend of a number of frequencies, thus forming a quasi-periodic feature in the power spectrum.
        
\begin{figure}[t]
            \centering
            \includegraphics[width=\hsize]{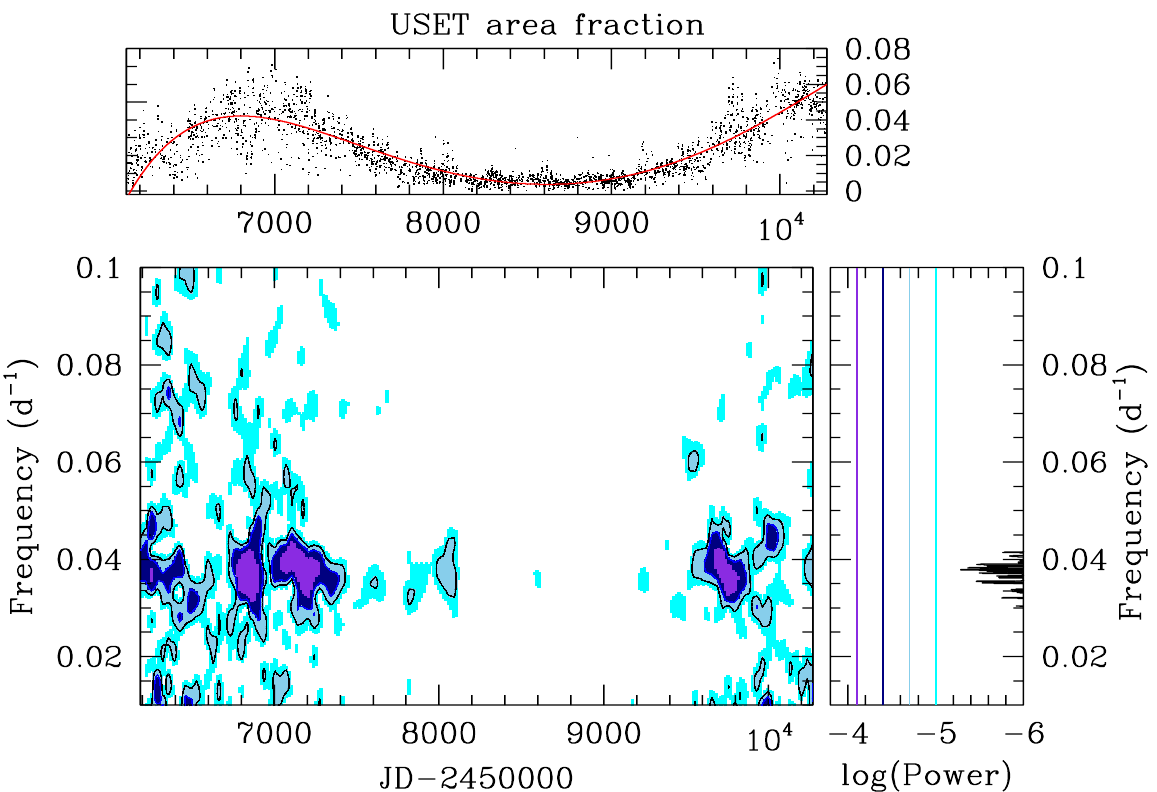}
            \caption{Time-frequency diagram of the $A_{PEN}$ time series. The top panel displays the observed data along with the red curve providing the fit by a polynomial to the long-term variations based on a degree of 6 . The color-scale image provides the evolution of the Fourier power spectrum with the epoch on the x-axis. Violet, dark blue, light blue, and cyan colors stand for power $\geq 8\,10^{-5}$, $\geq 4\,10^{-5}$, $\geq 2\,10^{-5}$, and $\geq 10^{-5}$. The right vertical panel illustrates the Fourier power spectrum evaluated over the full duration of the USET time series. The colored straight lines correspond to the scale used in the time-frequency diagram.}
            \label{fig:Time_frequency_diagram}
        \end{figure}

        An important question is whether the properties of this periodic modulation may vary over the activity cycle. Such variations can be expected, for instance, as a function of the overall activity level, as well as when the degree of uniformity of the longitudinal distribution of the plages and enhanced network changes. Indeed, a strictly uniform distribution would produce no modulation at all, regardless of the overall level of activity. To address this point, we have first de-trended the time series for the long-term (solar cycle) variations. To do so, we first splitted the total data set into 12 intervals of 347 days duration each. On each of these intervals, we evaluated the mean value of $A_{PEN}$ and adjusted the polynomial by a degree of 6 to these points. We tested lower degree polynomials, but 6 was the lowest degree that would allow us to represent the overall shape of the long-term variations well. This long-term trend is shown by the red curve in the top panel of Fig. \ref{fig:Time_frequency_diagram} and was then subtracted from the original USET time series. We then followed the approach of \cite{Rauw-2021} to build a time-frequency diagram, see Figure \ref{fig:Time_frequency_diagram}, by computing a Fourier transform of the data in sliding windows of 136 days duration, shifted in steps of 34 days\footnote{We also computed a time-frequency diagram with a sliding window of 680 days. Whilst this leads to a higher resolution in frequency, it degrades the temporal resolution, smearing out the temporal variations of the power spectrum. Still, the overall behavior is fully consistent with Fig. \ref{fig:Time_frequency_diagram}.}. 
        
        We can distinguish three different behaviors. For each case, we illustrate the link between the detection of a modulation due to the rotation and the associated magnetic structures present on the solar disk. For this purpose, we used synoptic maps showing the evolution of segmented bright chromospheric structures for several full solar rotations (see Figures in Appendix \ref{Annexe_A}, \ref{Annexe_B}, and \ref{Annexe_C}). They are constructed by juxtaposing the central part of consecutive deprojected segmented images. Those synoptic maps display the temporal evolution of the structures over several solar rotations (top panels) and the sum of bright pixels along vertical strips (bottom panels). These sums were evaluated in sliding windows with a width of 180$^{\circ}$ in longitude, corresponding to an entire daily image, and shifted in steps corresponding to the solar rotation over one day.

        First, the rotational modulation is most prominently seen during some time intervals near the maximum of cycle 24 and during one episode in the rising part of cycle 25. More specifically, there are three broad maxima of visibility of this modulation around JD2456800 (2014 May 22), JD2457200 (2015 June 26), and JD2459730 (2022 May 30). The first of these intervals lasts about 100 days, while the second and third one last about 200 days. Figures \ref{fig:20140522}, \ref{fig:20150626}, and \ref{fig:20220530}, referring to the broad maxima in Figure \ref{fig:Time_frequency_diagram}, show a succession of episodes with compact groups of plages and episodes with less activity. This is consistent with the results obtained in \cite{Dineva-2022},  for the same time interval, revealing the presence of rotation pattern separated into compact groups associated with large active regions rotating on and off the disk. The asymmetry in the longitudinal distributions could be related to the fact that some longitudes seem more favorable for the emergence of magnetic flux as observed for sunspot groups under the name of active longitudes (\citealp{Usoskin-2007}). Figures \ref{fig:20140522}, \ref{fig:20150626} and \ref{fig:20220530} further show that the modulation of the sum of bright pixels is not strictly periodic. Indeed, the synoptic maps indicate that the plages extend over a range in solar latitudes. Therefore, the observed modulation consists of the combination of signals having different rotational frequencies.
        
        Secondly, even near solar maximum there are time intervals during which there is no clear detection of rotational modulation. At those epochs, the distribution of the magnetic structures is nearly uniform in longitude as illustrated on the Figures \ref{fig:20131031} and \ref{fig:20141019}. Finally, as expected, the signal from the solar rotation is essentially absent during solar minimum because either no plage is present, or if there is one, it lasts for less than a rotation (see Figure \ref{fig:20200530}).

    \subsection{TIGRE S-index}\label{subsec:TIGRE_index_modulation}
        
\begin{figure}[t]
            \centering
            \includegraphics[width=0.49\hsize]{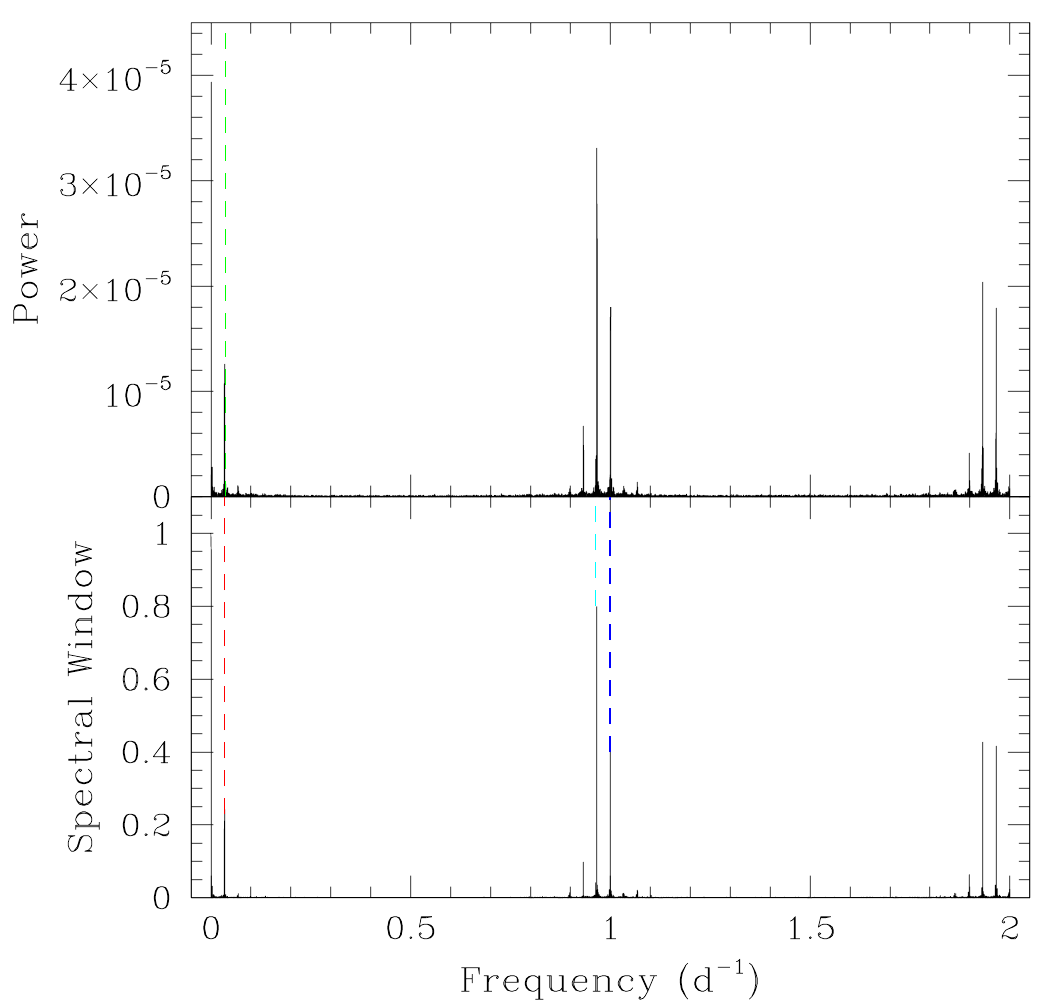}
            \includegraphics[width=0.49\hsize]{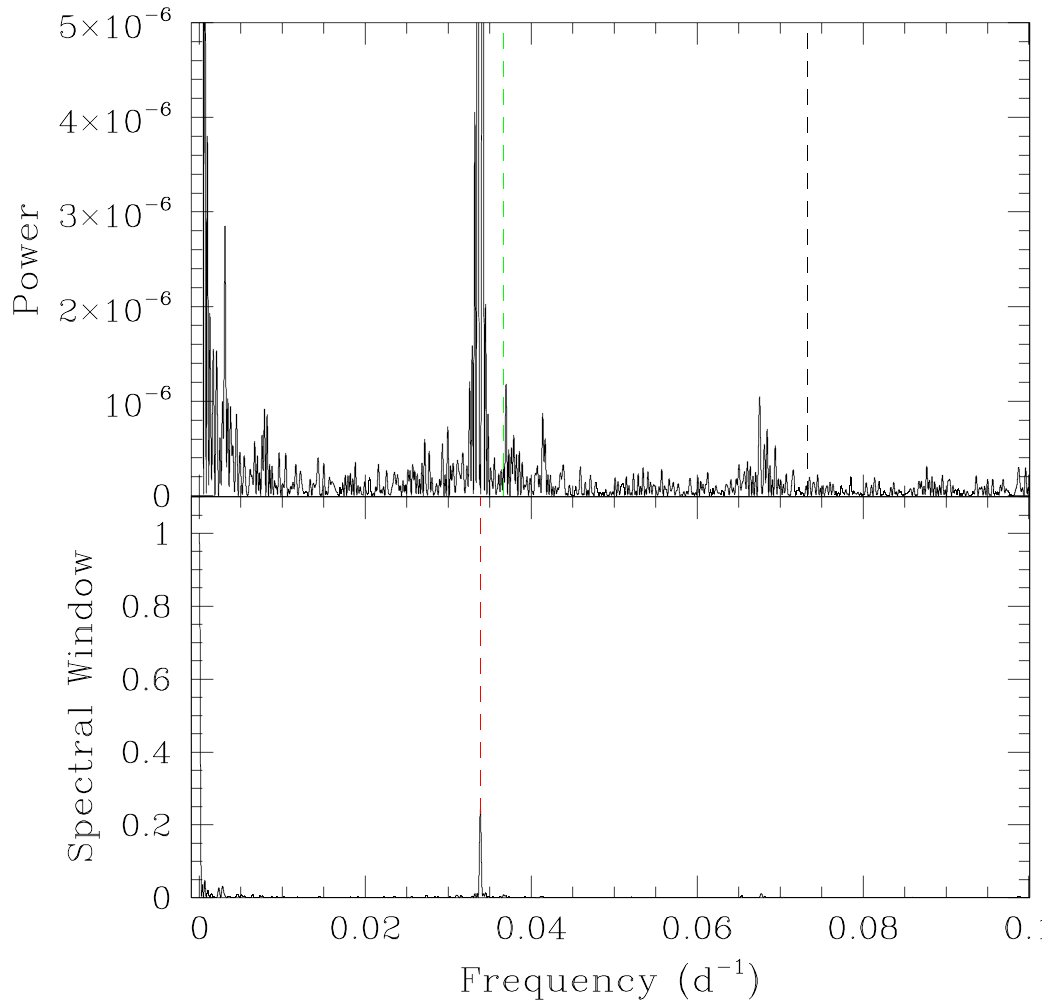}
            \caption{Fourier power spectrum (top-left) and spectral window (bottom-left) for the TIGRE solar S-index time series for frequencies between 0 and 2 d$^{-1}$ . The green dashed line in the power spectrum yields $\nu_{\rm Car}$, while the red, cyan and blue dashed lines in the spectral window plot identify the main aliasing frequencies which are respectively 0.0339 d$^{-1}$, 0.9661 d$^{-1}$, and 1.0000 d$^{-1}$. Right: Same details, but zooming on the frequency range between 0 and 0.1 d$^{-1}$. The dashed black line corresponds to $2 \nu_{\rm Car}$.}
            \label{fig:Fourier_TIGRE}
        \end{figure}

        As a consistency check, we also applied our Fourier method to the time series of the TIGRE S-index values. The spectral window unveils a more complex situation than for the USET data. Indeed, the main aliasing frequencies are found at 0.9661 d$^{-1}$, 1.0000 d$^{-1}$, and 0.0339 d$^{-1}$ (see Fig. \ref{fig:Fourier_TIGRE}). This latter aliasing frequency corresponds to one synodic month (29.53 days) and is due to the fact that TIGRE observes the Moon rather than the Sun. This also explains why the dominant aliasing frequency is found at 0.9661 d$^{-1}$ rather than 1.0000 d$^{-1}$: since the Moon rises  a bit later each night, the mean time interval between two consecutive observations is 1.035 days, rather than 1.000 day. Moreover, TIGRE cannot observe the Moon at phases too close to the New Moon; thus explaining the occurrence of the 0.0339 d$^{-1}$ peak in the spectral window (red dashed line). This situation has tremendous consequences on the power spectrum. Indeed, the fact that the Carrington synodic rotation period happens to be very close to the synodic month prevents us from obtaining a clear signature of a rotational modulation in the Fourier transform of the TIGRE S-index values. On the one hand, the aliasing via the 0.0339 d$^{-1}$ frequency replicates the low-frequency peaks which pollute the frequency domain near the Carrington frequency (see Fig. \ref{fig:Fourier_TIGRE}). On the other hand, the gaps in the time series of S-index measurements around each New Moon phase reduce the amplitude of the peaks associated with rotational modulation making them more difficult to distinguish against the overall noise level. This indicates that because of the TIGRE observing strategy, the S-index values are less sensitive to solar rotation than the USET data.
    
\section{Comparison between USET and TIGRE}\label{sec:USET_vs_TIGRE} 

   In this section, we study the correlation between $A_{PEN}$ and the solar S-index derived from TIGRE observations of the Moon. First, we characterize the temporal differences between both datasets (Section \ref{subsec:temporal_diff_tigre_uset}) and then study quantitatively their correlation (Section \ref{subsec:correlation_tigre_uset}).   

    \subsection{Temporal overlap between USET and TIGRE}\label{subsec:temporal_diff_tigre_uset}

         To compare the S-index and the area fraction, we first select overlapping data. This reduces the 2725 USET data to 790 images. Ideally the selected data should be recorded simultaneously. In practice, Brussels and Mexico are separated by seven time-zone hours but the fact that TIGRE observes the Moon and USET the Sun partially compensates the time difference. 
        
\begin{figure}[t]
            \centering
            \includegraphics[width=\hsize]{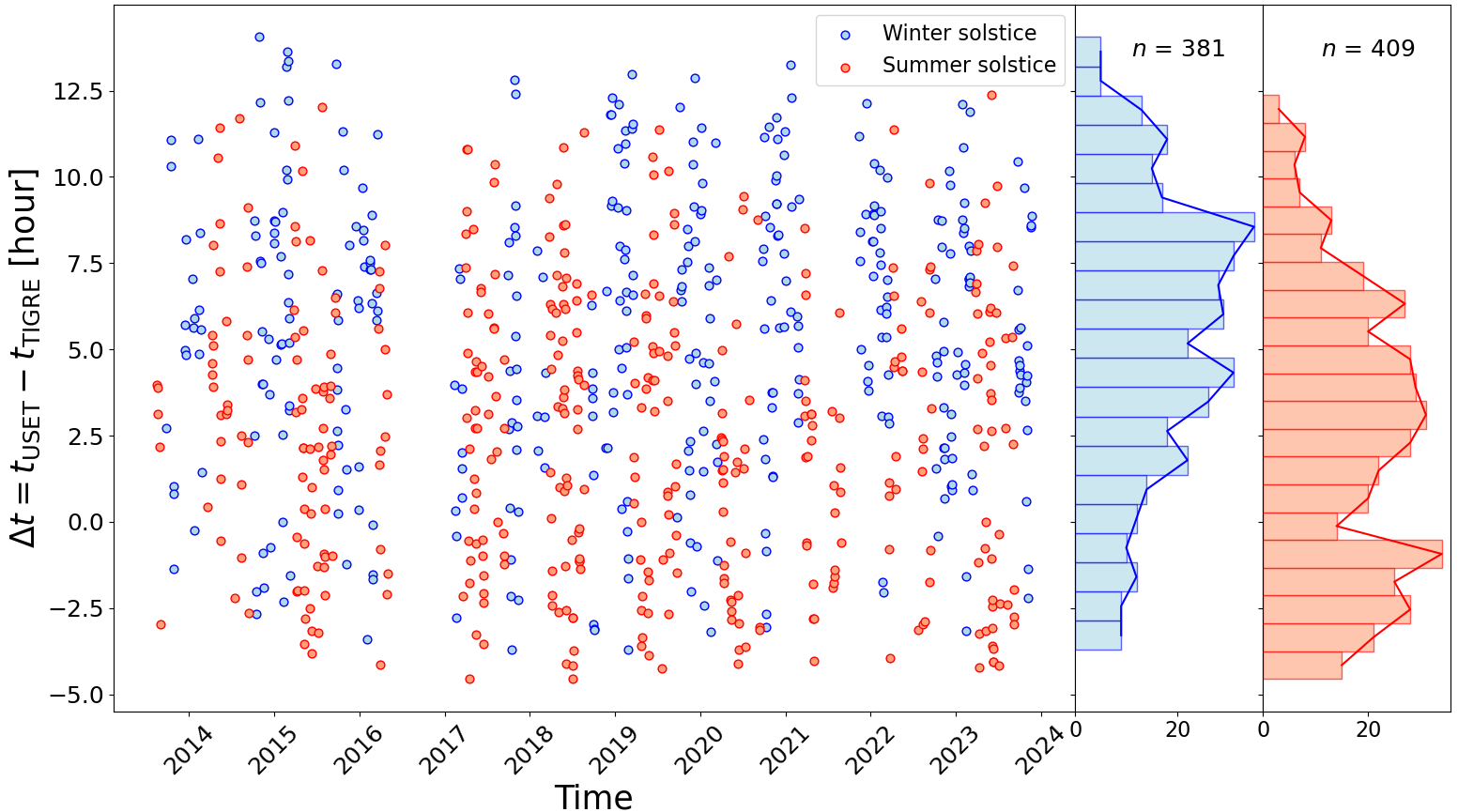}
            \caption{Time differences between USET and TIGRE. Data points are divided in two series: one that is six months around the winter solstice (in blue) and the other six months around the summer solstice (in red). Their corresponding histograms are displayed on the right panels.}
            \label{fig:Time_diff_USET_TIGRE}
        \end{figure}

        Figure \ref{fig:Time_diff_USET_TIGRE} shows the time differences between USET and TIGRE observations for the selected sample. Data are split in two groups based on the time of observation: six months around the winter solstice and six months around the summer solstice. We observe a seasonal effect on the time differences between USET and TIGRE: it is higher around the winter solstice than around the summer solstice. This is not surprising because in winter, the Sun rises later and sets earlier, so that in Brussels, we observe it a few hours later in the morning (observations start around 9-10 am); whereas, in Mexico, the night falls earlier so the observations are made earlier. The highest time differences correspond to situations where the USET observations in Brussels are made later in the afternoon, due to bad weather conditions in the morning for example and the TIGRE observations in Mexico are made earlier in the night. The absolute averaged time difference between USET and TIGRE is 4.13 hours and the impact of the appearance or disappearance of a structure is therefore negligible. Within this time, the Sun rotates by $\sim$ 2$^{\circ}$ and the variation of the area fraction is smaller than the uncertainties. Moreover, the limb structures are geometrically compressed, so their contribution to the total area fraction, $A_{PEN}$, is tiny.

    \subsection{Correlation between USET and TIGRE indices}\label{subsec:correlation_tigre_uset}
       
        \begin{figure}[t]
            \centering
            \includegraphics[width=0.9\hsize]{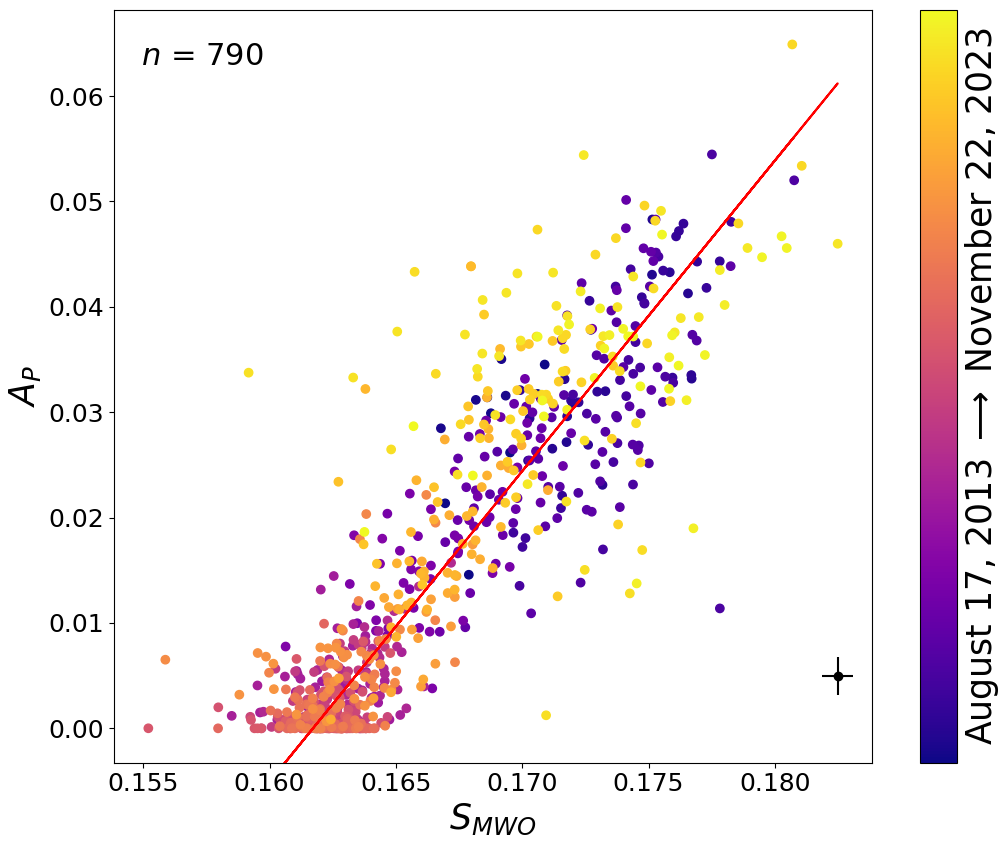}
            \includegraphics[width=0.9\hsize]{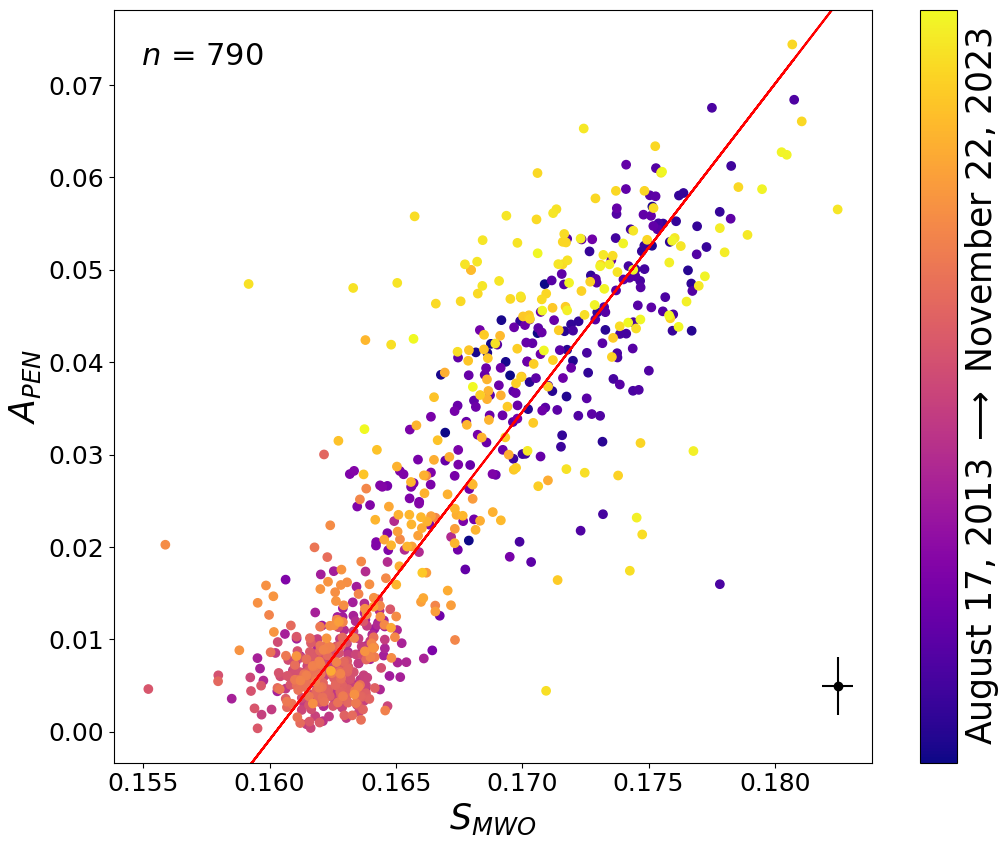}
            \caption{Correlation between daily values of the solar S-index from TIGRE (in the Mount Wilson scale) and the USET area fraction. Top panel: plages only, $A_P$. Bottom panel: plages and enhanced network, $A_{PEN}$. The parameter $n$ represents the number of data, and the chronological order is color-coded. The first and last date of the data are given next to the color bar. Additionally, a linear fit was performed to the data (red solid line). A mean error bar is displayed on the bottom right corner to have an idea of the uncertainties on the data.}
            \label{fig:USET_TIGRE}
        \end{figure}
        
        The correlation between the time overlapping area fraction from USET and the solar S-index from TIGRE is shown on Figure \ref{fig:USET_TIGRE}. The top panel displays the area fraction of the plages without considering the enhanced network, $A_P$, and the bottom panel represents the area fraction of the plages with the enhanced network, $A_{PEN}$. For plages area close to zero (top panel), there are still some decaying plages (which constitute the enhanced network) present on the solar surface and that contribute to the S-index. As expected, considering the enhanced network (bottom panel) makes a better correlation at low values and it reflects better the S-index. 
        
        First, we observe that the two datasets are linearly correlated, with a high Pearson correlation coefficient of 0.88, according to this equation:
        
        \begin{equation}
            \centering
            A_{PEN} = (3.55 \pm 0.06) \ S_{\textrm{MWO}} - (0.57 \pm 0.01)
            \label{eq:USET-TIGRE-equation}
        ,\end{equation} 
        where $A_{PEN}$ is the area fraction of the chromospheric structures (plages and enhanced network) from USET, and $S_{\textrm{MWO}}$ the solar S-index from TIGRE in the Mt. Wilson scale. 
                
        A second observation concerns the distribution of points around the fit. A small fraction of points (a bit more than 1\%) deviates with $3 \sigma$. This corresponds to images with a lower quality and where the segmentation is less robust. Apart from these outliers, there is a dispersion which is larger than the uncertainty on the data (displayed as a mean error on the bottom right corner). Various sources can explain this dispersion. One potential source is the image quality and the fact that residual non-radial inhomogeneity is still present, affecting the segmentation. Another hypothesis is that smaller and fainter bright structures not included in our disk-resolved index have a non-negligible contribution to the S-index. These elements could be the active network representing small bipolar areas and the quiet network defining the boundaries of the supergranulation \citep{2018-Meunier}.

\section{Discussion and conclusion}\label{sec:Discussion}

    We studied the correlation between the presence of bright structures in the solar chromosphere, based on USET images in the Ca {\sc ii} K line and the variability of the solar S-index, obtained with the TIGRE telescope on an overlapping period of ten years, spanning a large portion of cycle 24 to the beginning of cycle 25.

    We have constructed a disk-resolved time series from the USET images, segmenting the brightest structures which are the plages and the enhanced network. The correlation between the disk-resolved index and the S-index is well described by a linear relation, but a broad dispersion is present. Including small bright elements such as the active network and the quiet network could help to reduce the dispersion. Such a segmentation needs a more advanced processing of the image to remove the non-radial inhomogeneity produced by clouds and to compensate the turbulence, both effects being unavoidably present on images acquired by ground-based instruments. 

    For both time series, we have detected the modulation due to the rotation of bright structures on the disk and identified various behaviors at different phases of the solar cycle. From illustrative synoptic maps, we have shown that the detection of the modulation of the rotation is due to an asymmetry in the longitudinal distribution of the plages. This could be studied further and linked to the mechanism of active longitudes observed for sunspot groups. The detection of the rotation is intrinsically limited for TIGRE due to its observing strategy. In order to characterize the modulation on longer timescales, such as entire solar cycles, we would need long-term combined datasets. This could be done by taking advantage of existing long-term datasets in Ca {\sc ii} K and S-index. After a cross-calibration between the datasets, the correlation between these indices could then be studied on a longer period, covering the minimum of the solar cycle 23, stated as a low minimum \citep{Schroder-2012}.
    
    As the S-index is used to characterize the magnetic activity of Sun-like stars, a natural question arises as to whether stars seen under a different viewing angle, far from the equator, would also show the rotational modulation. We consider whether it is still possible to detect this rotational modulation for such inclinations and whether there is a specific inclination above which it is no longer possible to detect the modulation. Such a study could draw from on the present work, using the area fraction as a proxy for the S-index, and building an area fraction for different inclinations. This exploration will be the topic of a forthcoming paper.

\begin{acknowledgements}
      The authors wish to thank Theodosios Chatzistergos from the Max Planck Institute for Solar Research System, in Germany, for his help and his advice on the segmentation method. Grégory Vanden Broeck was supported by a PhD grant awarded by the Royal Observatory of Belgium. The USET instruments are built and operated with the financial support of the Solar-Terrestrial Center of Excellence (STCE).
\end{acknowledgements}

\bibliographystyle{aa}
\bibliography{biblio}

\begin{appendix}
    \onecolumn

\begin{landscape}

\section{Examples of synoptic map for the three broad maxima in Fig \ref{fig:Time_frequency_diagram}}\label{Annexe_A}
\raggedright The top panel of each figure display the evolution during several solar rotations of the segmented bright chromospheric structures (see Section \ref{subsec:USET_index}). The bottom panels show the sum of bright pixels of vertical strips (explained in Section \ref{subsec:USET_index_modulation}). The "CR" number above each synoptic map refers to the Carrington rotation number.

\vspace{-0.3cm}

\begin{figure}[h]
    \centering
    \includegraphics[scale=0.35]{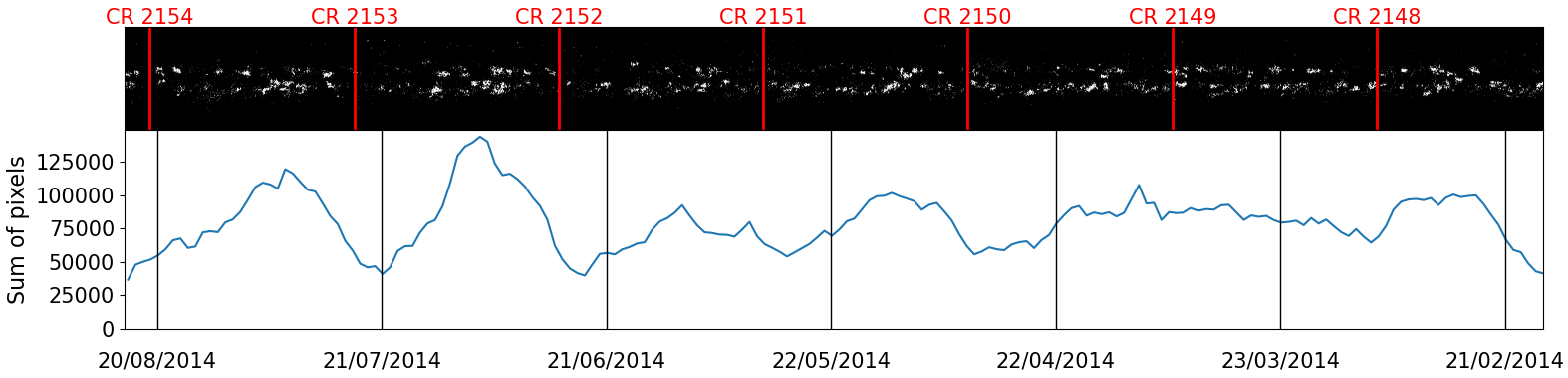}
    \vspace{-0.35cm}
    \caption{Synoptic map of the Sun centered around \textbf{May 22, 2014 (JD2456800)}}
    \label{fig:20140522}
    \vspace{0.5cm}
    \includegraphics[scale=0.35]{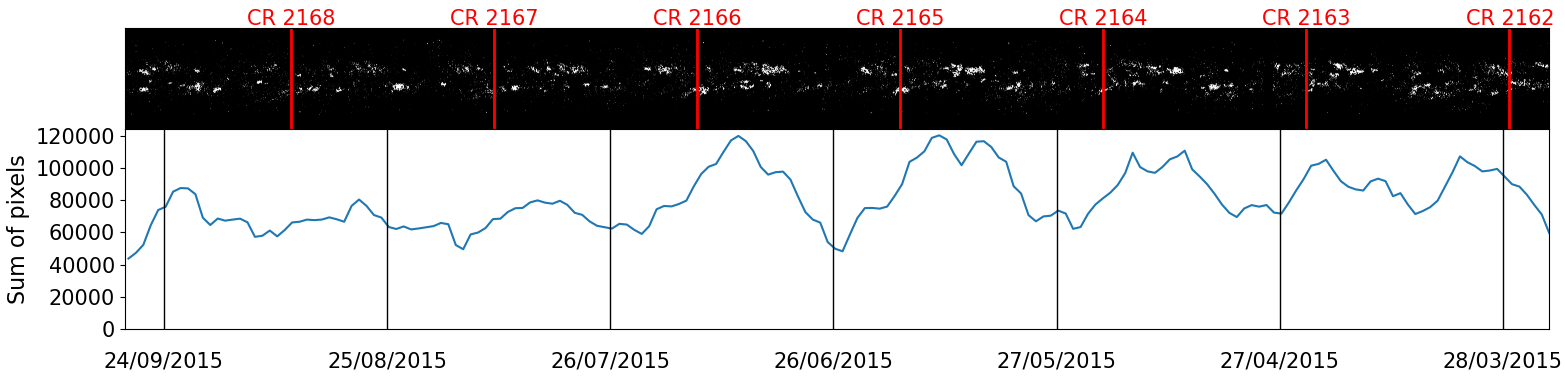}
    \vspace{-0.35cm}
    \caption{Synoptic map of the Sun centered around \textbf{June 26, 2015 (JD2457200)}}
    \label{fig:20150626}
    \vspace{0.5cm}
    \includegraphics[scale=0.35]{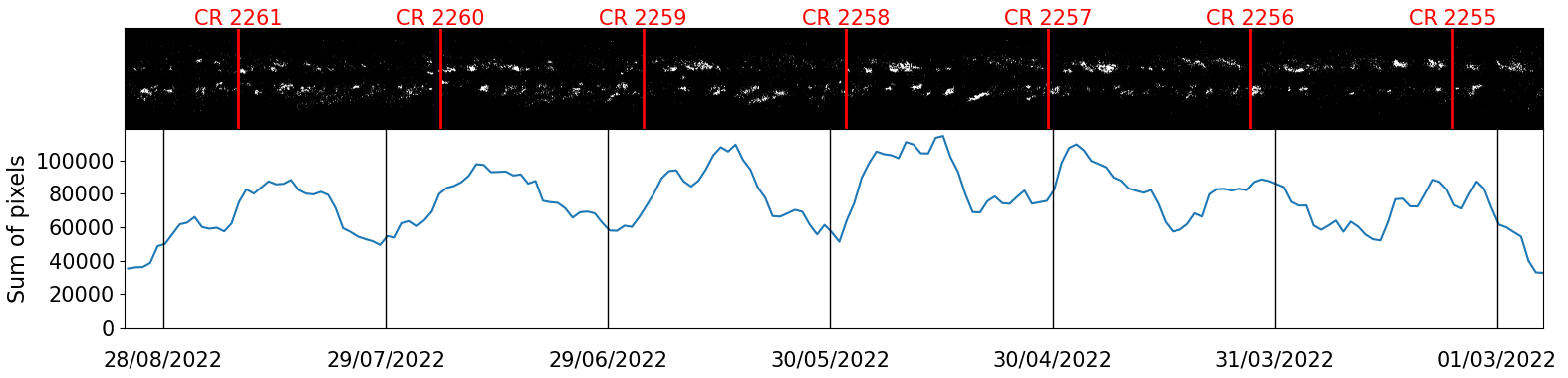}
    \vspace{-0.35cm}
    \caption{Synoptic map of the Sun centered around \textbf{May 30, 2022 (JD2459730)}}
    \label{fig:20220530}
\end{figure}  

\end{landscape}

    \onecolumn

\begin{landscape}

\section{Examples of synoptic map for episodes lacking detection of a modulation during solar maximum}\label{Annexe_B}
\raggedright The top panel of each figure display the evolution during several solar rotations of the segmented bright chromospheric structures (see Section \ref{subsec:USET_index}). The bottom panels show the sum of bright pixels of vertical strips (explained in Section \ref{subsec:USET_index_modulation}). The "CR" number above each synoptic map refers to the Carrington rotation number.

\vspace{-0.3cm}

\begin{figure}[h]
    \centering
    \includegraphics[scale=0.35]{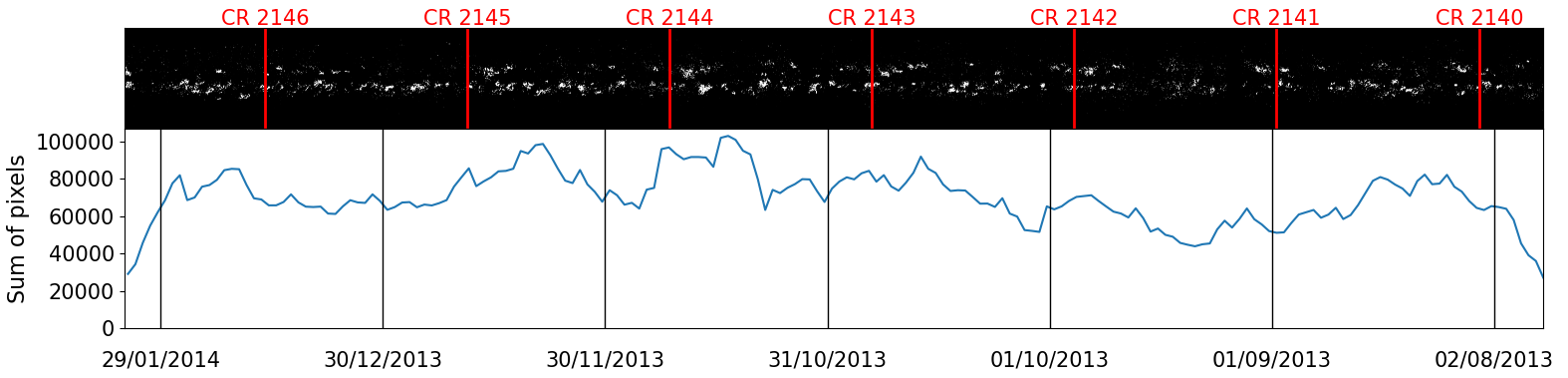}
    \vspace{-0.35cm}
    \caption{Synoptic map of the Sun centered around \textbf{October 31, 2013 (JD2456597)}}
    \label{fig:20131031}
    \vspace{0.5cm}
    \includegraphics[scale=0.35]{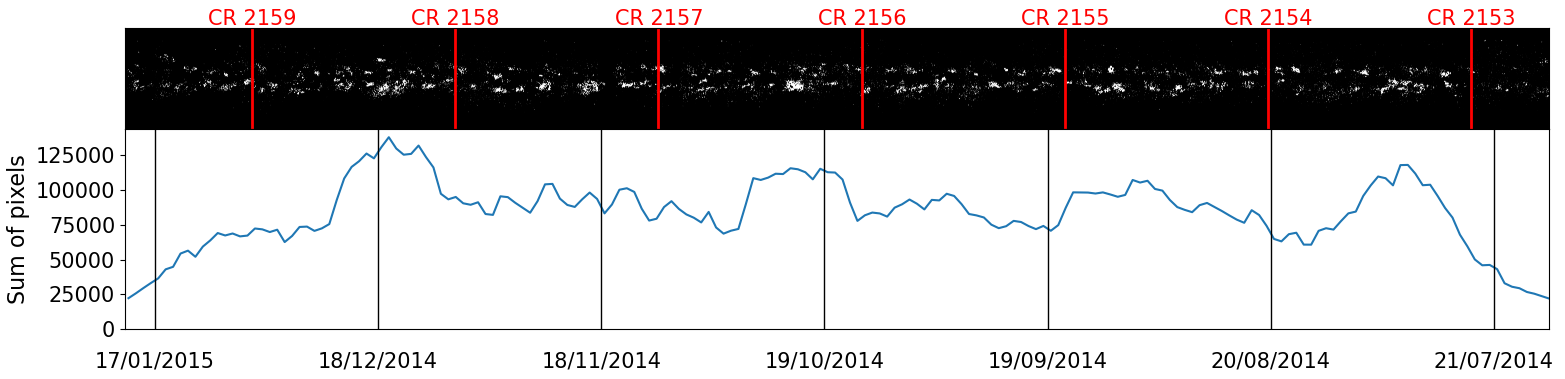}
    \vspace{-0.35cm}
    \caption{Synoptic map of the Sun centered around \textbf{October 19, 2014 (JD2456950)}}
    \label{fig:20141019}
\end{figure}

\end{landscape}

    \onecolumn

\begin{landscape}

\section{Example of synoptic map during the solar minimum}\label{Annexe_C}
\raggedright The top panel displays the evolution during several solar rotations of the segmented bright chromospheric structures (see Section \ref{subsec:USET_index}). The bottom panel shows the sum of bright pixels of vertical strips (explained in Section \ref{subsec:USET_index_modulation}). The "CR" number above each synoptic map refers to the Carrington rotation number.

\vspace{-0.3cm}

\begin{figure}[h]
    \centering
    \includegraphics[scale=0.35]{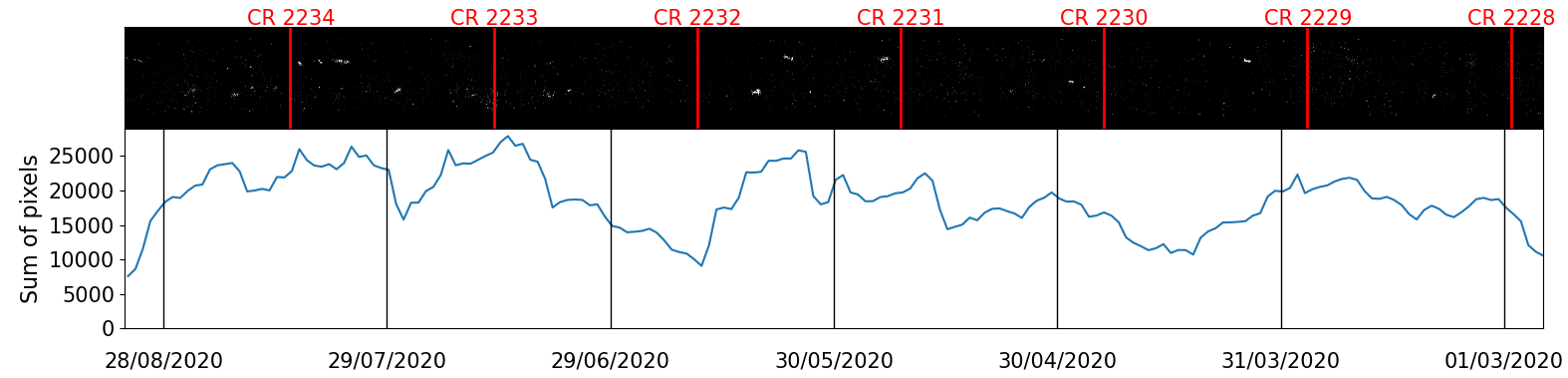}
    \vspace{-0.35cm}
    \caption{Synoptic map of the Sun centered around \textbf{May 30, 2020 (JD2459000)}}
    \label{fig:20200530}
\end{figure}

\end{landscape}

\end{appendix}

\end{document}